\documentclass[iop,revtex4]{emulateapj}
\usepackage{graphics,graphicx}
\usepackage{helvet}
\usepackage[utf8]{inputenc}
\usepackage[T1]{fontenc}
\usepackage{soul}
\usepackage{hyperref}
\usepackage{amsmath, amssymb, gensymb}
\usepackage{ulem}
\usepackage{natbib}
\usepackage{microtype}
\usepackage{multirow}
\usepackage{morefloats}
\usepackage{rotating}
\usepackage{xspace}
\usepackage{color}
\usepackage{xspace}
\def\Msol {$\hbox{M}_\odot$\xspace}

\def\mjybm {mJy\,beam$^{-1}$\xspace}
\def\ujybm {$\mu$Jy\,beam$^{-1}$\xspace}
\def\mjybmvert {$\left(\frac{\textrm{mJy}}{\textrm{beam}}\right)$\xspace}

\def\etal {\textit{et al.}\xspace}
\newcommand{\mir}[1]{\textbf{\fontfamily{lmvtt}\selectfont #1}} 

\begin{document}

\slugcomment{Submitted to ApJ on 16 Feb 2018; Accepted to ApJ on 16 Apr 2018}

\title{ALMA observations of polarization from dust scattering in the IM~Lup protoplanetary disk}
\shorttitle{ALMA observations of polarization from dust scattering in the IM~Lup protoplanetary disk}

\author{Charles L. H. Hull\altaffilmark{1,2,10}}
\author{Haifeng Yang\altaffilmark{3}}
\author{Zhi-Yun Li\altaffilmark{3}}
\author{Akimasa Kataoka\altaffilmark{4}}
\author{Ian W. Stephens\altaffilmark{5}}
\author{Sean Andrews\altaffilmark{5}}
\author{Xuening Bai\altaffilmark{6}}
\author{L. Ilsedore Cleeves\altaffilmark{5,11}}
\author{A. Meredith Hughes\altaffilmark{7}}
\author{Leslie Looney\altaffilmark{8}}
\author{Laura M. P\'erez\altaffilmark{9}}
\author{David Wilner\altaffilmark{5}}

\altaffiltext{1}{National Astronomical Observatory of Japan, NAOJ Chile Observatory, Alonso de C\'ordova 3788, Office 61B, 7630422, Vitacura, Santiago, Chile}
\altaffiltext{2}{Joint ALMA Observatory, Alonso de C\'ordova 3107, Vitacura, Santiago, Chile}
\altaffiltext{3}{Department of Astronomy, University of Virginia, Charlottesville, VA 22903, USA}
\altaffiltext{4}{National Astronomical Observatory of Japan, 2-21-1 Osawa, Mitaka, Tokyo 181-8588, Japan}
\altaffiltext{5}{Harvard-Smithsonian Center for Astrophysics, 60 Garden St., Cambridge, MA 02138, USA}
\altaffiltext{6}{Department of Physics, Tsinghua University, School of Sciences Building, Beijing 100084, China}
\altaffiltext{7}{Department of Astronomy, Wesleyan University, Van Vleck Observatory, 96 Foss Hill Drive, Middletown, CT 06457, USA}
\altaffiltext{8}{Department of Astronomy, University of Illinois, Urbana, IL 61801, USA}
\altaffiltext{9}{Departamento de Astronom\'ia, Universidad de Chile, Camino El Observatorio 1515, Las Condes, Regi\'on Metropolitana, Chile}
\altaffiltext{10}{NAOJ Fellow}
\altaffiltext{11}{Hubble Fellow}

\shortauthors{Hull \etal}
\email{chat.hull@nao.ac.jp}

\begin{abstract}
We present 870\,$\micron$ ALMA observations of polarized dust emission toward the Class II protoplanetary disk IM~Lup.  We find that the orientation of the polarized emission is along the minor axis of the disk, and that the value of the polarization fraction increases steadily toward the center of the disk, reaching a peak value of $\sim$\,1.1\%.  All of these characteristics are consistent with models of self-scattering of submillimeter-wave emission from an optically thin inclined disk.  The distribution of the polarization position angles across the disk reveals that while the average orientation is along the minor axis, the polarization orientations show a significant spread in angles; this can also be explained by models of pure scattering.  We compare the polarization with that of the Class I/II source HL~Tau.  A comparison of cuts of the polarization fraction across the major and minor axes of both sources reveals that IM~Lup has a substantially higher polarization fraction than HL~Tau toward the center of the disk.  This enhanced polarization fraction could be due a number of factors, including higher optical depth in HL~Tau, or scattering by larger dust grains in the more evolved IM~Lup disk.  However, models yield similar maximum grain sizes for both HL~Tau (72\,$\micron$) and IM~Lup (61\,$\micron$, this work).  This reveals continued tension between grain-size estimates from scattering models and from models of the dust emission spectrum, which find that the bulk of the (unpolarized) emission in disks is most likely due to millimeter (or even centimeter) sized grains. 
\\ 
\end{abstract}

\keywords{polarization --- scattering --- protoplanetary disks --- stars: formation --- stars: protostars}

\section{Introduction}
\label{sec:intro}

One of the longstanding goals of star- and disk-formation enthusiasts has been to make a well resolved map of the magnetic field in a protoplanetary disk.  At the $\gtrsim$\,100\,au scales of protostellar envelopes, the assumption to date has been that polarized emission from thermal dust grains is perpendicular to the magnetic field, where spinning dust grains are aligned via radiative torques with their short axes parallel to the magnetic field; see, e.g., \citet{Lazarian2007} and \citet{Andersson2015}.  If polarization detected toward a disk were indeed produced by magnetically aligned grains, it would provide the long-sought-after evidence that young protostellar disks are magnetized; this is a prerequisite for the operation of the magneto-rotational instability (MRI; \citealt{Balbus1991}) and magnetized disk winds \citep{Blandford1982, Turner2014}, both of which are thought to play a crucial role in disk evolution.

Over the last two decades, polarimetric observations with the Berkeley-Illinois-Maryland Array (BIMA), the Submillimeter Array (SMA), the Combined Array for Research in Millimeter-wave Astronomy (CARMA), the mid-infrared polarimeter CanariCam on the Gran Telescopio Canarias (GTC), and now the Atacama Large Millimeter/submillimeter Array (ALMA) have progressively achieved higher resolution and sensitivity, enabling observations of polarized thermal dust emission on scales of protostellar cores, envelopes, and now protoplanetary disks.  The SMA, CARMA, and ALMA mapped the inferred magnetic field toward large numbers of low-mass protostellar cores with $\sim$\,100--1000\,au resolution  \citep[e.g.,][]{Girart2006, Rao2009, Hull2013, Stephens2013, Hull2014, Hull2017a, Hull2017b, Cox2018}, and a few studies with the GTC, SMA, CARMA, and Karl G. Jansky Very Large Array (VLA) searched for (and were sometimes able to marginally resolve) what was assumed to be the magnetic field structure at $\lesssim$\,100\,au scales in a few protostellar and protoplanetary disks \citep{Hughes2009b, Hughes2013, Rao2014, Stephens2014, Cox2015, SeguraCox2015, DLi2016, Liu2016, FernandezLopez2016, DLi2018}.

Just as ALMA was poised to finally achieve the goal of making resolved images of magnetic fields in protoplanetary disks, several theoretical studies predicted that polarized (sub)millimeter-wave emission from disks could be produced partially---or completely---by the self-scattering of dust emission from (sub)millimeter-sized grains in those disks \citep{Kataoka2015, Kataoka2016, Pohl2016, Yang2016a, Yang2016b}, following on previous work by \citet{Cho2007}.  \citet{Kataoka2016} and \citet{Yang2016b} used this mechanism to explain the 1.3\,mm CARMA polarization observations of the Class I/II source HL~Tau by \citet{Stephens2014} and the polarization pattern observed in 870\,$\micron$ ALMA observations of the transition disk HD 142527 \citep{Kataoka2016b}.  More recent results include observations by \citet{CFLee2018} of the edge-on HH 212 and HH 111 disks, whose polarization can be interpreted as arising from either scattering or magnetically aligned grains; and observations by \citet{Girart2018} of the massive HH 80-81 disk, whose polarization may arise from dust self-scattering and/or alignment with an anisotropic radiation field (see below).

A further layer of complexity was introduced when \citet{Tazaki2017} proposed a third mechanism that can produce polarization in disks: namely, the alignment of dust grains with their short axes parallel to the (radial) dust emission gradient (this work was based on the radiative torque model by \citealt{Lazarian2007}).  This mechanism, sometimes referred to as ``radiative alignment,'' is distinct from dust self-scattering but, like self-scattering, acts independently of the disk's magnetic field.  Polarization from this mechanism, which has an azimuthal morphology, appears consistent with 3\,mm ALMA polarization observations of the HL~Tau disk \citep{Kataoka2017}.

The differing polarization morphologies at long (3\,mm) versus short (870\,$\micron$) wavelengths in various disks suggested that multi-wavelength observations of the same source would be essential in order to disentangle the contributions of polarization due to dust self-scattering, alignment with the dust emission gradient, and magnetically aligned dust grains.  This goal was first achieved by \citet{Kataoka2017}, who compared  3\,mm ALMA observations of HL~Tau with previous 870\,$\micron$ (SMA) and 1.3\,mm (CARMA) observations by \citet{Stephens2014}. The polarization morphology at the shorter wavelengths observed by CARMA and the SMA is well explained by dust self-scattering \citep{Kataoka2016, Yang2016a}, which, in an inclined disk like HL~Tau, manifests itself as polarization aligned with the minor axis of the disk.  However, the longer wavelength (3\,mm) observations can be explained by alignment with the dust emission gradient \citep{Tazaki2017}. This was clearly confirmed by \citet{Stephens2017b}, who presented well resolved ALMA observations of polarization toward HL~Tau at 870\,$\micron$ and 1.3\,mm, in addition to the 3\,mm data reported in \citet{Kataoka2017}.  The polarization morphologies at each wavelength were dramatically different: the 870\,$\micron$ map showed clear evidence of dust self-scattering, and the 1.3\,mm data showed a roughly equal superposition of the patterns from self-scattering and from alignment with the dust emission gradient.

To shed more light on the origins of dust polarization in disks, and to investigate the polarization in a Class II source that is more evolved than HL~Tau, we performed 870\,$\micron$ ALMA observations of the disk surrounding the Class II source IM~Lup.  \citet{Finkenzeller1987} and \citet{Martin1994} classified IM~Lup as a weak-line T Tauri star, based on its relatively narrow $H_\alpha$ line and lack of optical veiling. It is sometimes considered a transition object between a classical and a weak-line T Tauri star \citep{Pinte2008}.  Recent measurements of the mass accretion rate suggest a value of $10^{-8}$\,$M_\odot$\,yr$^{-1}$ \citep{Alcala2017}, typical for T Tauri stars.\footnote{Previous observations by \citet{Gunther2010} suggested an extremely low accretion rate on the order of $10^{-11}$\,$M_\odot$\,yr$^{-1}$.  An accretion rate this low would imply a magnetic field strength of less than $\sim$\,100\,$\mu$G on the 100\,au scale \citep{Bai2009}, which would be too weak to align grains $\gtrsim$\,1\,$\micron$ in size \citep{Hughes2009b}.  However, the difference between this value and the value reported by \citet{Alcala2017} could be due to variability in the accretion rate.} There is evidence for substantial grain growth in the IM~Lup disk---up to millimeter sizes---based on detailed modeling of multi-frequency data \citep{Pinte2008}. These millimeter-sized grains, if present, would be even less aligned by the (weak) magnetic field; these same large grains would also increase the scattering cross section, resulting in a brighter polarization signal.

In addition, IM~Lup has a number of other features that make it ideally suited for studying scattering-induced disk polarization. These include a disk that is both massive ($\sim$\,0.1\,\Msol) and large ($\sim$\,$4\arcsec$, or 600\,au in diameter), the absence of a contaminating envelope, azimuthal symmetry, and an intermediate disk inclination ($i\approx 48^\circ$, where $i=0^\circ$ for face-on).  IM~Lup has been studied extensively at millimeter wavelengths \citep[e.g.,][]{Lommen2007, Pinte2008, Panic2009, Cleeves2016, Tripathi2017}.  All but the inner $\lesssim$\,40\,AU of the disk is optically thin at 1.3\,mm and 870\,$\micron$ \citep{Cleeves2016}.  

Below we discuss our ALMA observations in \S\,\ref{sec:obs} and results in \S\,\ref{sec:res}.  In \S\,\ref{sec:dis} we discuss a number of issues, including a comparison of the polarization fraction in IM~Lup and HL~Tau (\S\,\ref{sec:pfrac_comparison}); grain growth, and a maximum grain-size estimate for IM~Lup (\S\,\ref{sec:grain_size}); dust settling and optical depth (\S\,\ref{sec:dust_settling}); and disk magnetic fields (\S\,\ref{sec:disk_Bfields}).  We offer our conclusions and potential paths forward in \S\,\ref{sec:con}.

\section{Observations}
\label{sec:obs}

We used ALMA to observe dust polarization at 870\,\micron{} toward IM~Lup on 2017 April 24 and 2017 April 26.  The pointing center was $\alpha_\textrm{J2000}$ = 15:56:09.172, $\delta_\textrm{J2000}$ = --37:56:06.483.  The observations have a synthesized beam (resolution element) of 0$\farcs$50 $\times$ 0$\farcs$40 at a position angle of 76.9$\degree$, 
corresponding to a linear resolution of $\sim$\,72\,au at a distance of $161 \pm 10$\,pc \citep{Gaia2016}.  

The largest recoverable scale in the data is approximately 4$\arcsec$, which matches the largest extent (i.e., the diameter) of the dust emission in IM~Lup's disk (the CO emission extends beyond a diameter of $\sim$\,$12\arcsec$, or $\sim$\,2000\,au; \citealt{Panic2009, Cleeves2016}).  The ALMA polarization data comprise 8\,GHz of wide-band dust continuum ranging in frequency from $\sim$\,336.5--350.5\,GHz, with a mean frequency of 343.479\,GHz (873\,$\micron$).  The flux, bandpass, and phase calibrators were Titan, J1517-2422, and J1610-3958, respectively; these calibrators were chosen automatically by querying the ALMA source catalog when the project was executed.  The polarization calibrator, J1549+0237, was chosen by hand because of its high polarization fraction. At Band 7 (870\,\micron{}), ALMA's flux-calibration accuracy is $\sim$\,10\%, as determined by the observatory flux monitoring program.  For a detailed discussion of the ALMA polarization system, see \citet{Nagai2016}.  Note that the uncertainties quoted in this work are all statistical unless otherwise specified.

The dust continuum images were produced by using the Common Astronomy Software Applications (CASA, \citealt{McMullin2007}) task \mir{TCLEAN} with a Briggs weighting parameter of robust\,=\,0.5.  The images were improved by performing four rounds of phase-only self-calibration using the total intensity (Stokes $I$) image as a model (the shortest interval for determining the gain solutions was 10\,s).  The Stokes $I$, $Q$, and $U$ maps (where the $Q$ and $U$ maps show the polarized emission) were each cleaned independently with an appropriate number of \mir{TCLEAN} iterations after the final round of self-calibration.  The rms noise level in the final Stokes $I$ dust map is $\sigma_I = 100$\,\ujybm{}, whereas the rms noise level in the map of polarized intensity $P$ (see Equation \ref{eqn:P} below) is $\sigma_P = 22$\,\ujybm{}.  The reason for this difference in noise levels is that the total intensity image is more dynamic-range limited than the polarized intensity images.  

The quantities that can be derived from the polarization maps are the polarized intensity $P$, the fractional polarization $P_\textrm{frac}$, and the polarization position angle $\chi$:

\begin{align}
P &= \sqrt{Q^2 + U^2} \label{eqn:P} \\
P_\textrm{frac} &= \frac{P}{I} \\
\chi &= \frac{1}{2} \arctan{\left(\frac{U}{Q}\right)}\, .
\end{align}

\noindent
Note that $P$ has a positive bias because it is always a positive quantity, even though the Stokes parameters $Q$ and $U$ from which $P$ is derived can be either positive or negative.  This bias has a particularly significant effect in low-signal-to-noise ($< 5\,\sigma$) measurements.  Note that while we do debias the polarized intensity map as described in \citet{Vaillancourt2006} and \citet{Hull2015b}, it has only a very minor effect on our results, as $\sim$\,80\% of the polarization detections we report have a signal-to-noise ratio $> 5\,\sigma_P$.  

See Table \ref{table:data} for the ALMA polarization data, which includes $I$; $P$; $\chi$ and its uncertainty $\delta\chi$; and $P_{\textrm{frac}}$ and its uncertainty $\delta P_{\textrm{frac}}$; at every position where polarization was detected (see the top panel of Figure \ref{fig:obs}).  The systematic uncertainty in linear polarization observations with ALMA is 0.03\% (corresponding to a minimum detectible polarization of 0.1\%).

\section{Results}
\label{sec:res}

\begin{figure*}
\begin{center}
\includegraphics[width=0.8\textwidth, clip, trim=0cm -1cm 2cm 0.3cm]{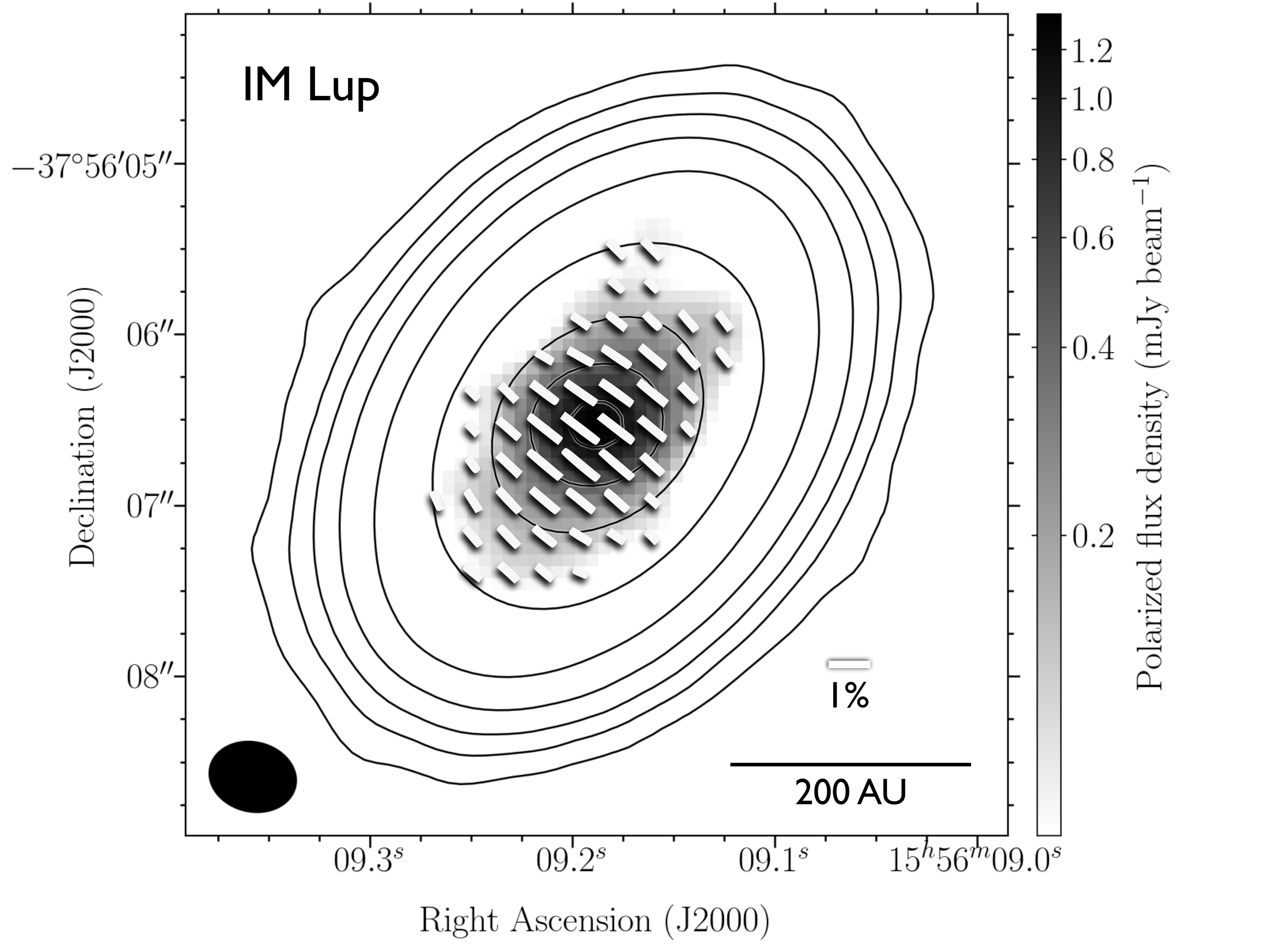}
\includegraphics[width=1.0\textwidth, clip, trim=0cm 10cm 0cm 0cm]{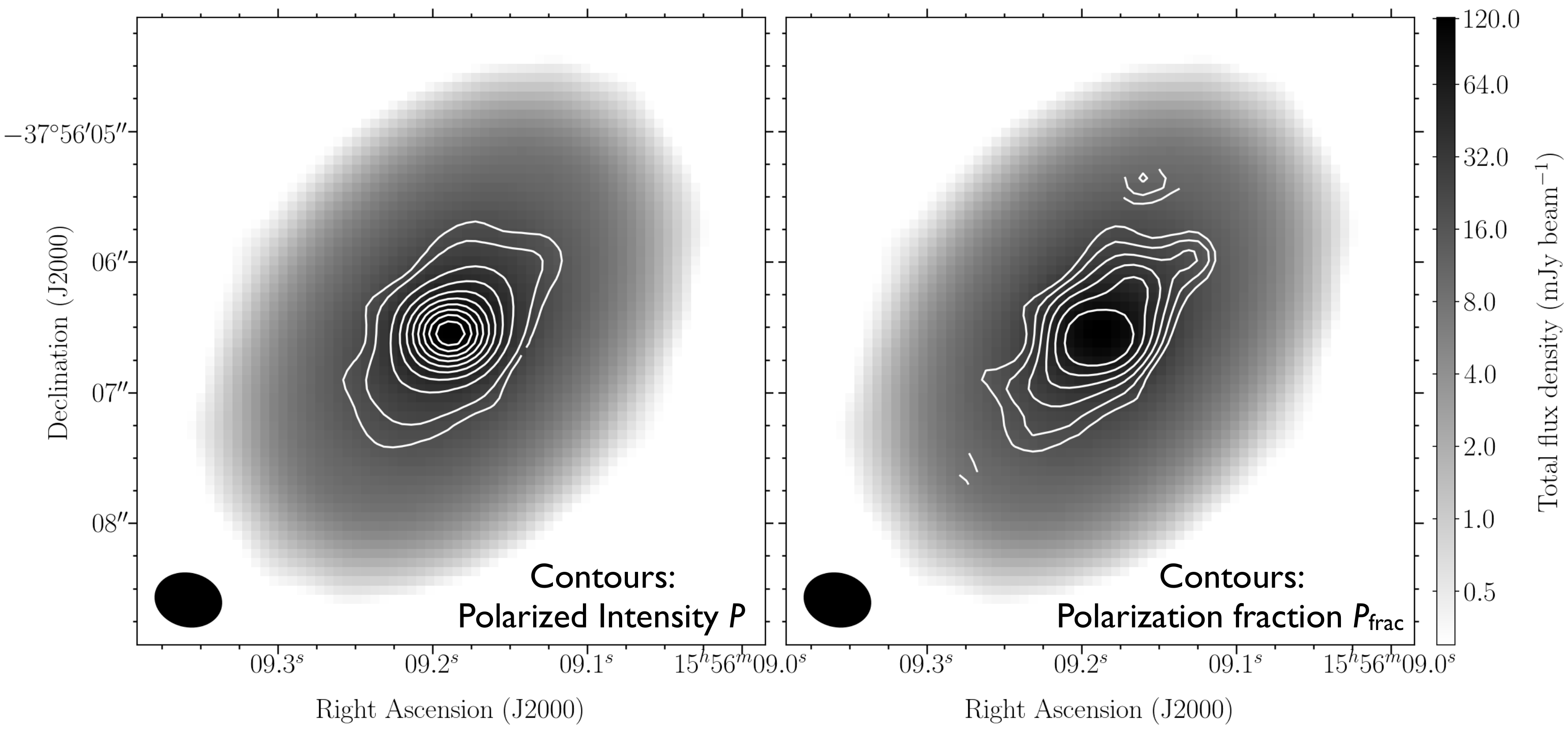}
\end{center}
\vspace{-1em}
\caption{\footnotesize
870\,$\micron$ ALMA maps of the Class II protoplanetary disk IM~Lup.  The peak values of the total intensity (Stokes $I$), polarized intensity ($P$), and polarization fraction images are 120.74\,\mjybm, 1.37\,\mjybm, and 0.011, respectively.   The rms noise values in the total intensity and polarized intensity thermal dust emission images are $\sigma_I = 100$\,\ujybm and $\sigma_P = 22$\,\ujybm, respectively.  The black ellipses in the lower-left corners of all panels represent the ALMA synthesized beam (resolution element), which measures 0$\farcs$50 $\times$ 0$\farcs$40 at a position angle of 76.9$\degree$, corresponding to a linear resolution of $\sim$\,72\,au at a distance of $161 \pm 10$\,pc \citep{Gaia2016}.  
\textit{Top:}  Polarization map of IM~Lup.  Contours are the total intensity thermal dust emission, plotted at 3,\,8,\,16,\,32,\,64,\,128,\,256,\,512,\,1024\,$\times$\,$\sigma_I$.  Grayscale is the polarized thermal dust emission, plotted starting at 3\,$\sigma_P$.  Line segments are the polarization orientation $\chi$ of the dust emission, with lengths proportional to the polarization fraction $P_{\textrm{frac}}$.
\textit{Bottom left:} Grayscale is the total intensity thermal dust emission; contours are the polarized intensity $P$, plotted at 0.06,\,0.1,\,0.2,\,0.3,\,0.4,\,0.5,\,0.6,\,0.7,\,0.8,\,0.9\,$\times$ the peak of 1.37\,\mjybm.
\textit{Bottom right:} Grayscale is the total intensity thermal dust emission; contours are the polarization fraction $P_{\textrm{frac}} = P/I$, plotted at levels of 0.005,\,0.006,\,0.007,\,0.008,\,0.009,\,0.01.  The coordinates and the grayscale are identical in the two bottom panels.
\textit{The ALMA data used to make this figure are available in the online version of this publication.}
\vspace{1em}
}
\label{fig:obs}
\end{figure*}

The results of our full-polarization, 870\,\micron{} dust continuum observations toward the IM~Lup protoplanetary disk are shown in Figure \ref{fig:obs}.  The large, upper panel depicts the polarized intensity $P$ (grayscale), total intensity $I$ (contours), and the orientation of the polarized emission (white line segments).  The polarization is resolved in $\sim$\,13 independent beams across the bright, central region of the disk.  When Nyquist sampled (plotted twice per synthesized beam in each dimension), this yields 52 polarization orientations, which are plotted everywhere where there is significant polarized emission, i.e., they have not been masked based on the Stokes $I$ map.\footnote{In the case of IM~Lup, all polarization detections happen to be coincident with significant Stokes $I$ emission.  However, this is not always the case: see, for example, the ALMA polarization observations of Serpens SMM1 in \citealt{Hull2017b}, where the authors detect highly significant polarized dust emission in regions where there is no significant Stokes $I$.}  We also plot contour maps of the polarized intensity $P$ (Figure \ref{fig:obs}, lower-left panel) and the polarization fraction $P_\textrm{frac}$ (lower-right panel), superposed on grayscale images of the total intensity.  The integrated 870\,\micron{} flux of IM~Lup from our observations is 580\,mJy, consistent with previous ALMA observations at the same frequency \citep{Cleeves2016}.

To test for frequency dependence in the polarization toward IM~Lup, we made images of polarized intensity and polarization angle from each of the four individual correlator bands in the dataset.  As mentioned in \S\,\ref{sec:obs}, these bands ranged in frequency from $\sim$\,336.5--350.5\,GHz; the maximum difference of 14\,GHz yields a fractional bandwidth difference of $\sim$\,4\%.  We found no significant changes in polarized intensity, polarization fraction, or polarization angle as a function of frequency, indicating that multi-wavelength studies of disk polarization (e.g., the work on HL~Tau by \citealt{Stephens2017b}) will require observations at multiple distinct ALMA bands with wide frequency separations.

As has been the case with several polarization results from ALMA including \citet{Stephens2017b} and \citet{Vlemmings2017}, we detect a marginal circularly-polarized signal in the Stokes $V$ map; however, the circular polarization fraction is only $\sim$\,0.17\% of the total intensity.  This value is smaller than the current 0.6\% systematic uncertainty in ALMA circular polarization observations, and thus could be spurious.

\subsection{Distribution of polarization angles across the IM~Lup disk}

\begin{figure}
\begin{center}
\includegraphics[width=0.5\textwidth, clip, trim=0cm 0cm 0cm 0cm]{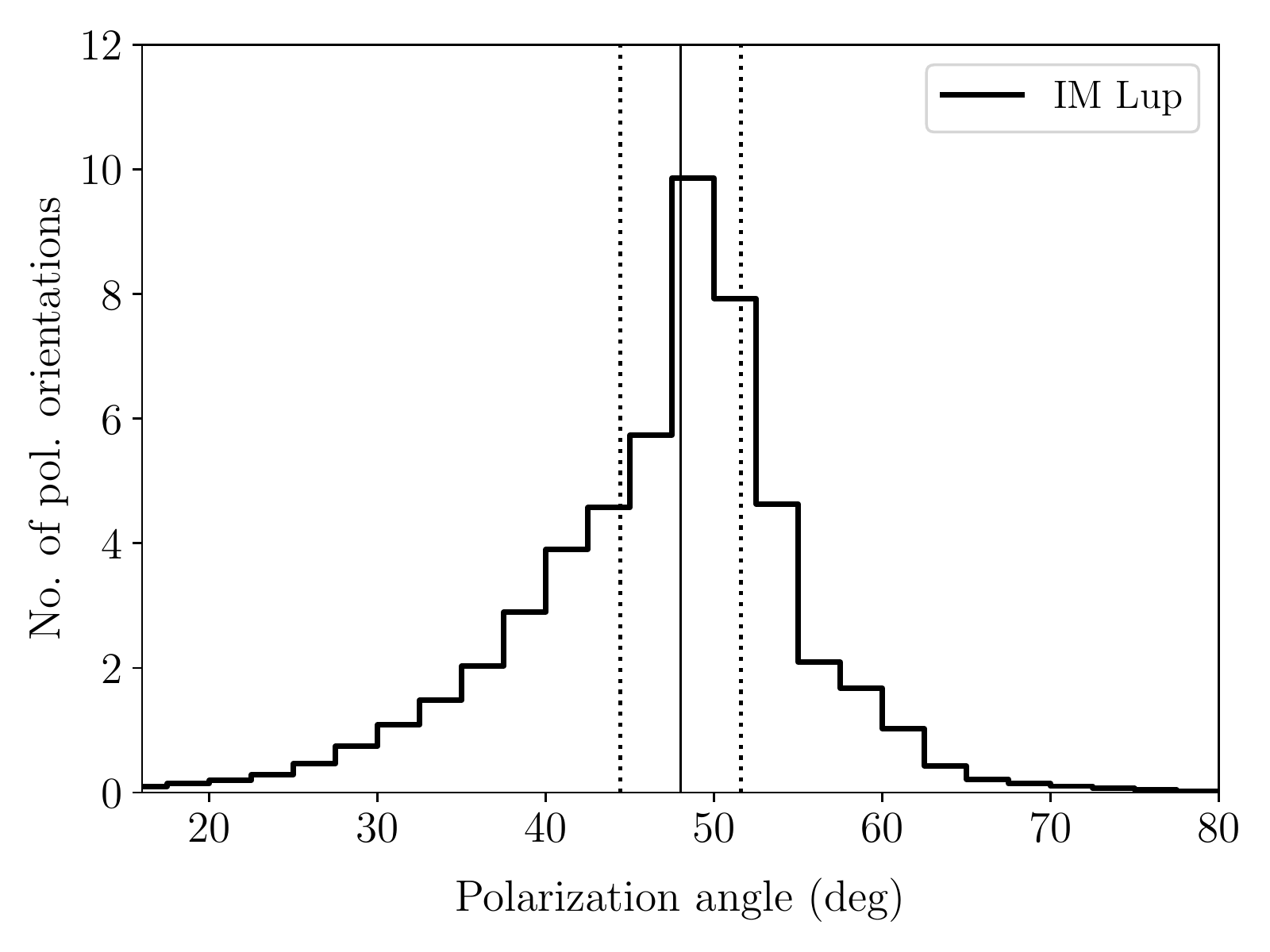}
\end{center}
\vspace{-1em}
\caption{\footnotesize
Histogram of the polarization angles across the map of IM~Lup seen in Figure \ref{fig:obs}, where each individual polarization angle measurement was binned presuming it had a Gaussian distribution with the mean and standard deviation listed in Table \ref{table:data}.
The solid vertical line is the 48$\degree$ orientation of the minor axis of IM~Lup's disk.  The two dotted vertical lines are plotted at $ 48 \pm 3.6\degree$, where the 3.6$\degree$ uncertainty is the mean of the statistical uncertainties in all of the polarization angles detected toward IM~Lup.  
The broad distribution of polarization angles indicates that many of the polarization angles deviate significantly from the 48$\degree$ orientation of the minor axis of the disk.
\vspace{1em}
}
\label{fig:hist}
\end{figure}

In Figure \ref{fig:hist} we plot a histogram of the polarization orientations in the map of IM~Lup shown in Figure \ref{fig:obs}, where each individual polarization angle measurement was binned presuming it had a Gaussian probability distribution function (PDF) with the mean and standard deviation listed in Table \ref{table:data}.  
The peak of the histogram depends on the number of bins, but it is always consistent with the $\sim$\,48$\degree$ orientation of the minor axis of the 
disk
to within the $\pm$\,3.6$\degree$ uncertainty plotted in the histogram.  3.6$\degree$ is the mean of the statistical uncertainties in all of the polarization angles detected toward IM~Lup.  We choose this mean value to represent the intrinsic statistical scatter in the polarization angles because each angle has an uncertainty that is dependent on the signal-to-noise of the polarized intensity at the given location in the disk (see Table \ref{table:data}).

A notable feature of the distribution of polarization angles in Figure \ref{fig:hist} is the significant width beyond the bounds of the uncertainty, revealing that many of the polarization angles deviate significantly from the orientation of IM~Lup's minor axis.  This is the result of a slight azimuthal curvature in the polarization orientations furthest from the disk center (along the major axis), seen in the top panel of Figure \ref{fig:obs}.  This is consistent with the pure-scattering models by \citet{Yang2016a, Yang2017}, as well as with the morphology of the polarization in the 870\,$\micron$ observations of HL~Tau reported in \citet{Stephens2017b}.  Note, however, that while the slight azimuthal curvature can be explained by pure self-scattering, it could also be due to the superposition of polarized emission both from self-scattering and from dust grains aligned with the dust emission gradient \citep{Tazaki2017}.  This superposition can be seen most clearly in the 1.3\,mm ALMA image of HL~Tau reported in \citet{Stephens2017b}; however, those data show far more curvature in the polarization angles than the 870\,$\micron$ IM~Lup data we present here, suggesting that alignment of dust grains with the dust emission gradient is not contributing significantly to our 870\,$\micron$ polarization observations.

\section{Discussion}
\label{sec:dis}

\subsection{Comparing the polarization fraction profiles of IM~Lup and HL~Tau}
\label{sec:pfrac_comparison}

These observations of IM~Lup are the second 850\,$\micron$ ALMA polarization observations of an inclined, full (e.g., non-transition) disk, after the HL~Tau observations published in \citet{Stephens2017b}.  Both the IM~Lup and the HL~Tau observations show characteristics that are broadly consistent with models of dust self-scattering from an inclined disk.  

\begin{figure*}
\begin{center}
\includegraphics[width=0.49\textwidth, clip, trim=0cm 0cm 0cm 0cm]{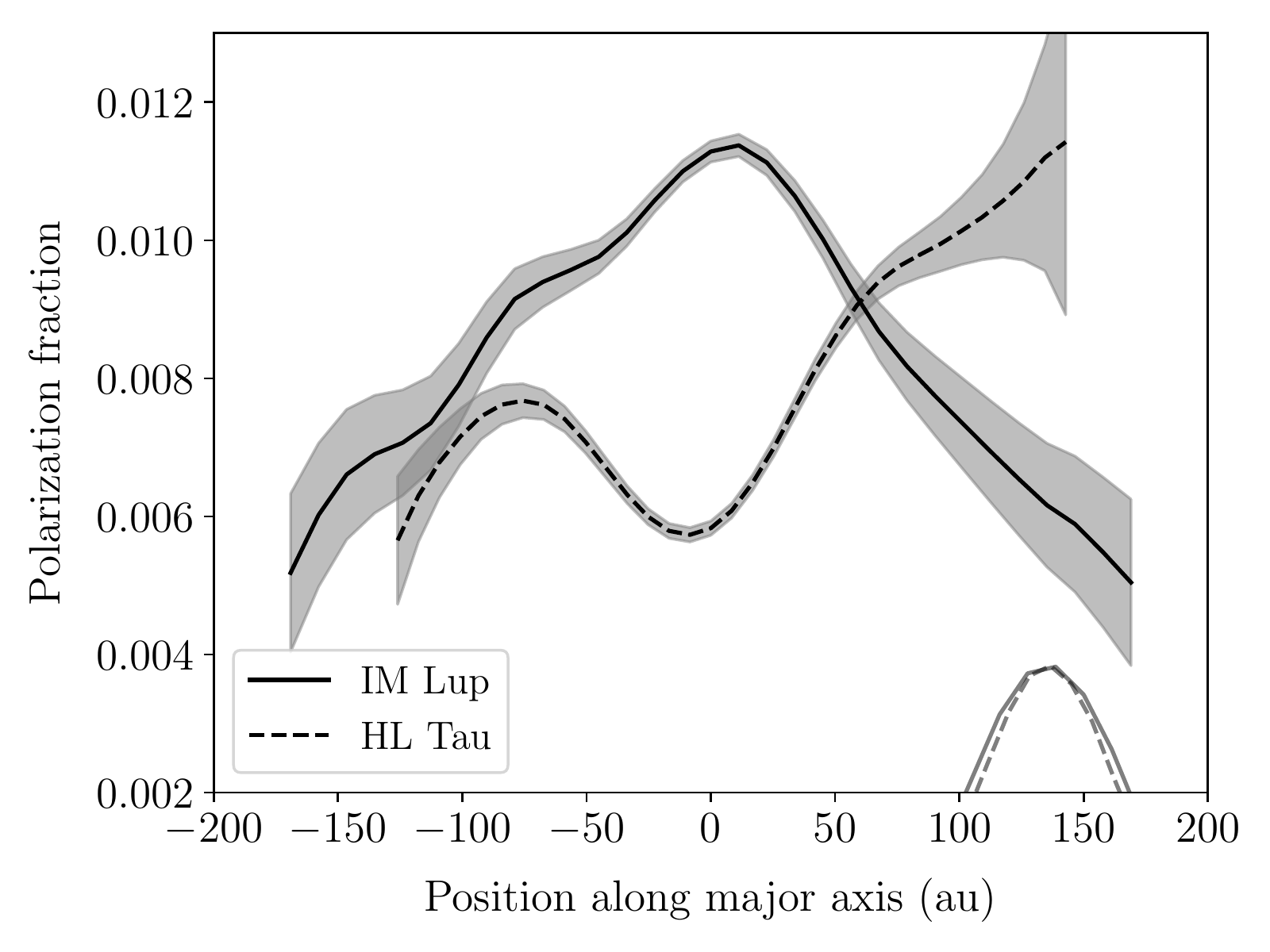}
\includegraphics[width=0.49\textwidth, clip, trim=0cm 0cm 0cm 0cm]{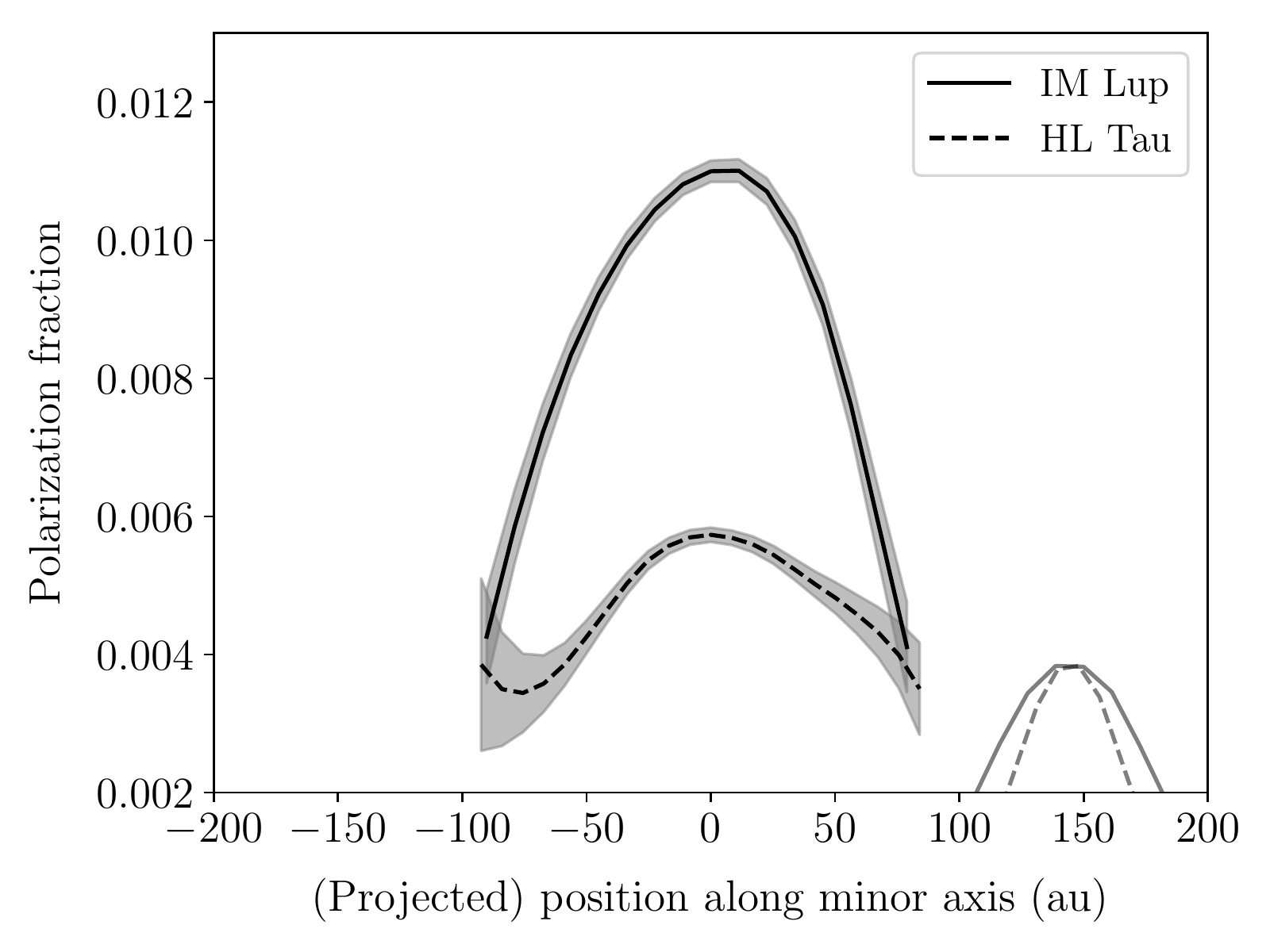}
\end{center}
\vspace{-1em}
\caption{\footnotesize
1-dimensional cuts of the polarization fraction $P_\textrm{frac}$ across the major axis (left) and minor axis (right) of both IM~Lup (this work) and the 870\,$\micron$ HL~Tau data from \citet{Stephens2017b}.  The minor axis has not been deprojected.  The gray shaded regions have a vertical width of $\pm$\,$\delta P_{\textrm{frac}}$ (see Table \ref{table:data} for the IM~Lup data); in the highest signal-to-noise regions, these statistical uncertainties can be smaller than the 0.03\% systematic uncertainty in ALMA polarization observations.  
The gray curves in the lower-right are cuts of the synthesized beams; the plot's horizontal axis lies at the beams' half-maximum levels.  
The spatial resolutions of the IM~Lup and HL~Tau observations are approximately 72\,au and 55\,au, respectively.
\vspace{1em}
}
\label{fig:cuts}
\end{figure*}

However, substantial differences between the two sources are revealed in Figure \ref{fig:cuts}, which shows 1-dimensional cuts of the polarization fraction $P_\textrm{frac}$ across the major axis (left) and minor axis (right) of both IM~Lup and HL~Tau (the 870\,$\micron$ HL~Tau data are from \citealt{Stephens2017b}).  Both the major- and minor-axis profiles of IM~Lup peak at the center.  However, only HL~Tau's minor-axis profile peaks at the center; its major-axis profile has a central depression in the polarization fraction.  As shown in \citet{Stephens2017b}, the polarization orientations at 870\,$\micron$ are roughly along the minor axis, thus it is unlikely that the depression is because of plane-of-sky smearing of the polarization signal within the central synthesized beam (as opposed to the 3\,mm ALMA map of HL~Tau shown in \citealt{Kataoka2017} and \citealt{Stephens2017b}, where the central beam is smeared as a result of the azimuthal polarization pattern).  It is more likely that the depression along the major axis is caused by HL~Tau's high optical depth at 870\,$\micron$ -- 1.3\,mm wavelengths \citep{ALMA2015, CarrascoGonzalez2016, Jin2016}, which is predicted to reduce the polarization fraction \citep{Kataoka2015, Yang2017}.  

Alternatively, the fact that the polarization fraction in IM~Lup is substantially higher than in HL~Tau could be because IM~Lup is a more evolved source than HL~Tau, and thus has had a longer time for grains to grow larger.  However, considering the similar estimates for the maximum grain size in the two sources (see \S\,\ref{sec:grain_size}), it is not clear that this is the case. 


\subsection{Maximum grain-size estimate for IM~Lup}
\label{sec:grain_size}

To estimate the maximum size of the dust grains responsible for the observed scattering polarization, we adopt the IM~Lup disk model based on \citet{Cleeves2016}, taking into account only millimeter-sized dust grains. The model is one of a viscous disk \citep{LyndenBellPringle1974} with a column density of dust $\Sigma_c = 0.25\,\rm g\,cm^{-2}$ at a radius $R_c = 100\rm\,au$. The scale height of the millimeter grains is $H_c = 3$\,au at $R_c$, which is $0.25$\,$\times$ the scale height of the gas and the micron-sized grains. The surface-density power-law index $\gamma = 1.0$. For the dust grain model, we adopted the one used by \citet{Kataoka2015} and \citet{Yang2016a}, which assumes spherical grains comprising a mixture of water ice (62\%), organics (30\%), and silicates (8\%).  All fractional abundances are by volume and are taken from \citet{Pollack1994}. The grain size distribution follows the Mathis, Rumpl \& Nordsieck (MRN)-type power-law $n(a) \propto a^{-3.5} $ \citep{Mathis1977}, with a fixed minimum grain size $a_{\rm min}=0.25\rm\, \mu m$. 

We calculate the absorption/scattering opacities, as well as the full scattering phase matrix, with Mie theory \citep{Bohren1983}.  We then calculate polarization by varying the maximum grain size $a_{\rm max}$ with the Monte Carlo Radiative Transfer code RADMC-3D \citep{Dullemond2012}.  We can reproduce the observed polarization in our 870\,$\micron$ IM~Lup observations with $a_{\rm max}$ of 61\,$\micron$.  This is very close to the 72\,$\micron$ value estimated by \citet{Yang2016a}, who modeled 1.3\,mm HL~Tau observations by \citet{Stephens2014}.

\begin{figure*}
\begin{center}
\includegraphics[width=0.5\textwidth, clip, trim=1cm 0cm 0.3cm 0cm]{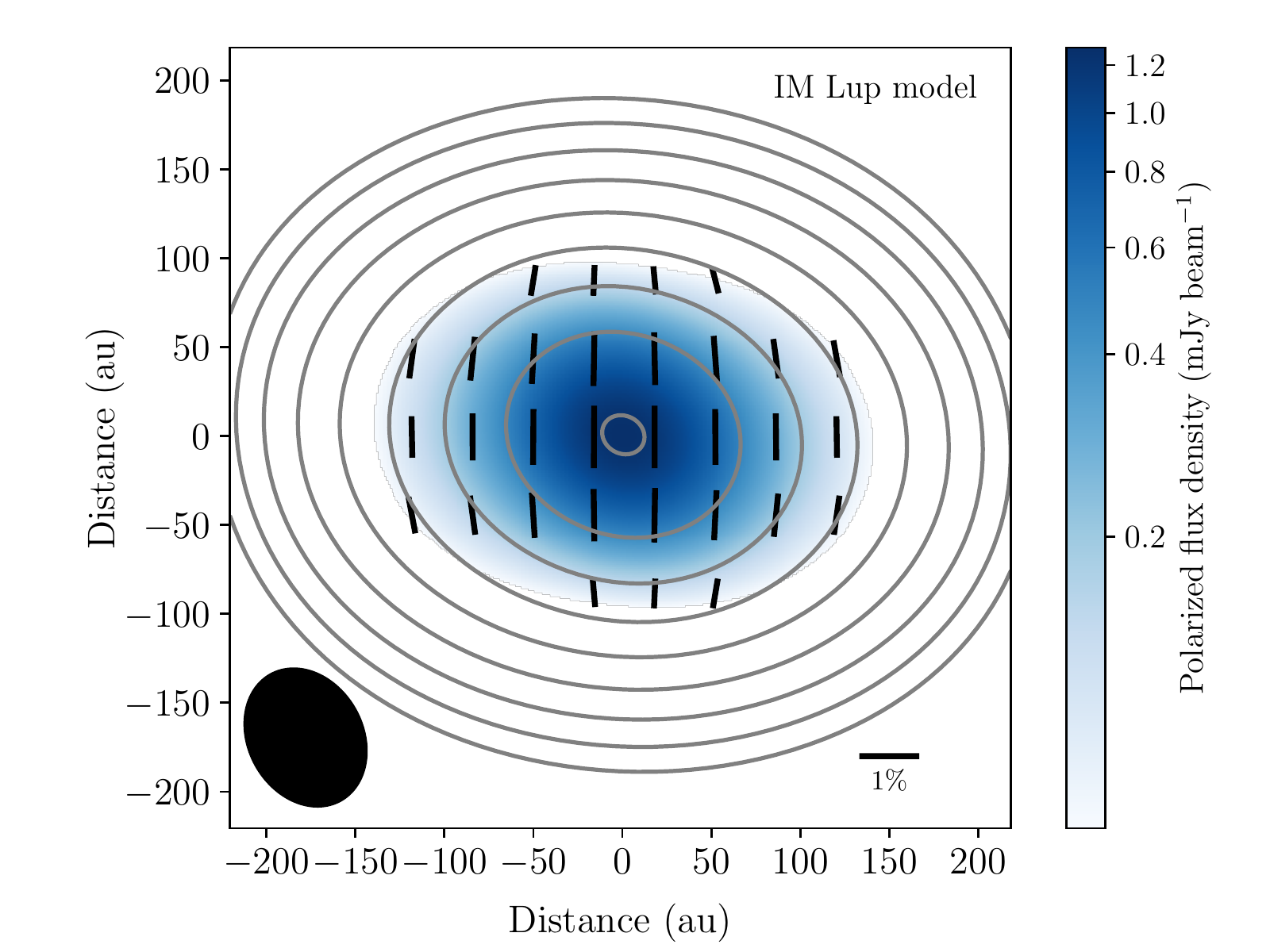}
\includegraphics[width=0.48\textwidth, clip, trim=0cm 0cm 0cm 0cm]{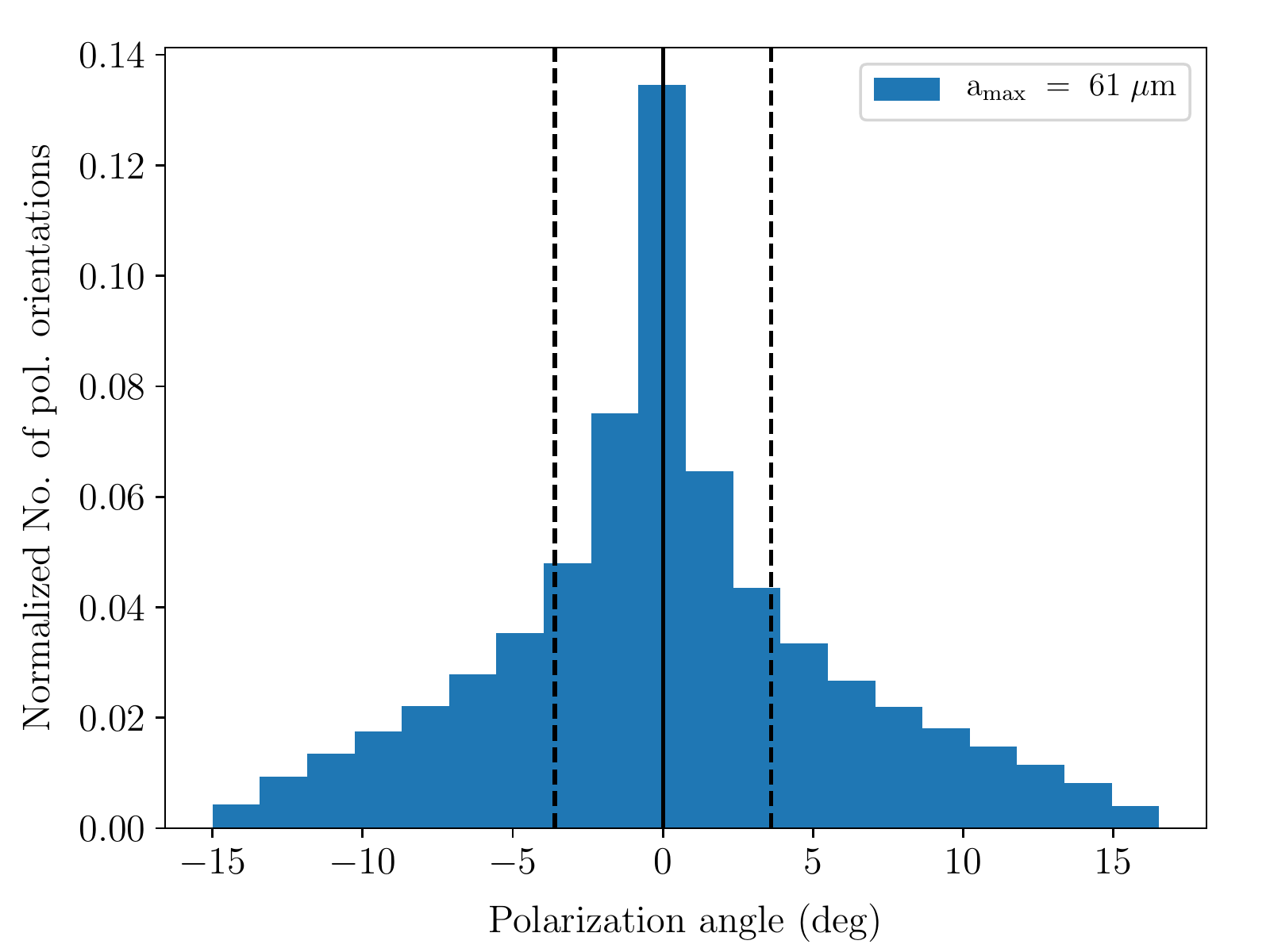}
\includegraphics[width=0.49\textwidth, clip, trim=0cm 0cm 0cm 0cm]{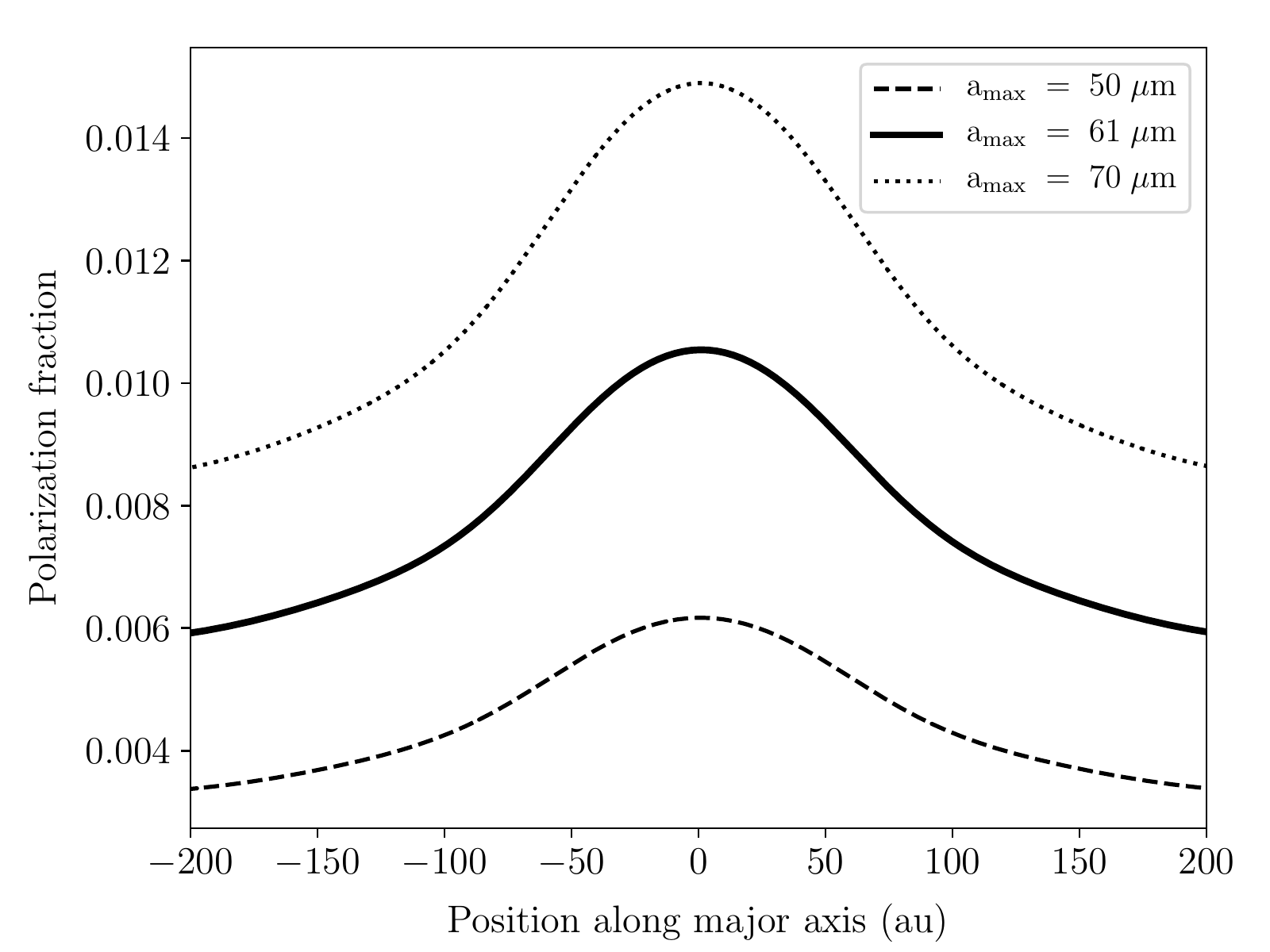}
\includegraphics[width=0.49\textwidth, clip, trim=0cm 0cm 0cm 0cm]{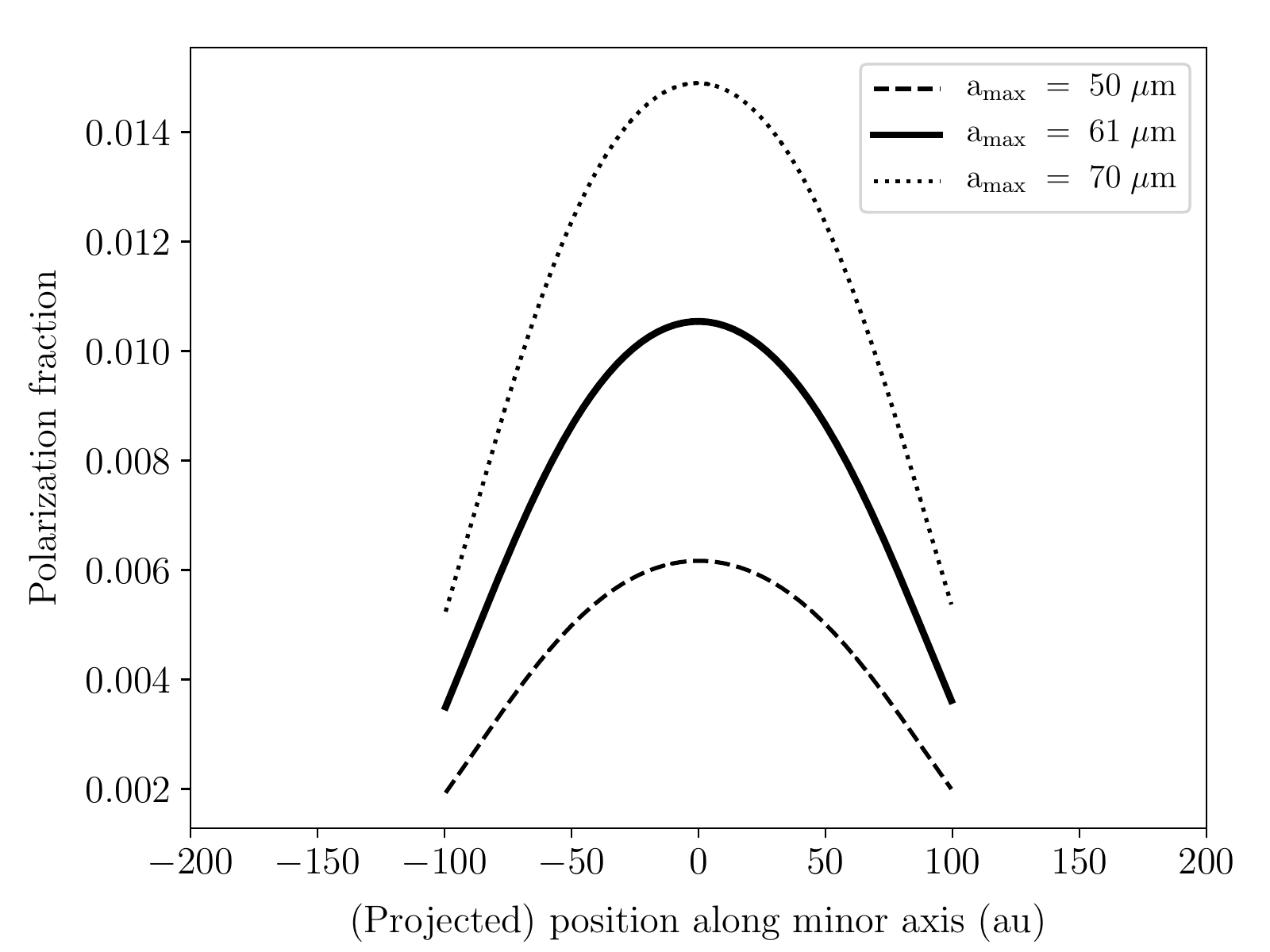}
\end{center}
\vspace{-1em}
\caption{\footnotesize
Best-fit model to the ALMA 870\,$\micron$ observations of IM Lup.  
\textit{Upper-left:} The contours represent the total intensity (Stokes $I$), which are logarithmically scaled beginning at 3\,$\sigma_I$ = 300\,$\mu$Jy\,beam$^{-1}$. The color scale represents the polarized intensity, and is also logarithmically scaled with a minimum value of 3\,$\sigma_P$ = 66\,$\mu$Jy\,beam$^{-1}$.  The line segments represent the polarization orientation.
\textit{Upper-right:} Normalized histogram of polarization angles in the model.  0$\degree$ corresponds to the minor axis orientation. The two dashed lines are $\pm 3.6\,\degree$, the same as Figure \ref{fig:hist}.
\textit{Lower-left and lower-right:} The polarization fraction along the major and (projected) minor axes, respectively, in the models.  The thick, solid line reflects the best-fit model with a maximum grain size of 61\,$\micron$; the dotted and dashed lines reflect models with larger (70\,$\micron$) and smaller (50\,$\micron$) maximum grain sizes, which do not fit the data as well.
\vspace{1em}
}
\label{fig:model}
\end{figure*}

We show the results of our best-fit model in Figure \ref{fig:model}.  We find that the histogram of the polarization angles in the model (Figure \ref{fig:model}, upper-right panel) is in good agreement with the that of the observations (Figure \ref{fig:hist}), in that both histograms show a significant spread in angles. In addition, the predicted distributions of the polarization fraction along the major and minor axes agree well with observations, peaking at $\sim$\,1.1\% when the maximum grain size is set to be 61\,$\micron$ (see the thick, solid lines in the bottom panels of Figure \ref{fig:model}).  We tested slightly larger (70\,$\micron$) and smaller (50\,$\micron$) maximum grain sizes, and found that they do not reproduce the observations (see the dotted and dashed curves in the bottom panels of Figure \ref{fig:model}).

As a consistency check, we repeat the exercise of \citet{Kataoka2017}, who analyzed the polarization of HL~Tau using a metric they refer to as the ``total polarization fraction,'' which is simply the integrated polarized intensity $P$ divided by the integrated total intensity $I$.  We find that the total polarization fraction of IM~Lup in our 870\,$\micron$ observations is $P/I \approx 0.50\%$; for the HL~Tau observations from \citet{Stephens2017b}, the total polarization fractions are 0.62\% (870\,$\micron$) and 0.67\% (1.3\,mm).  Estimating the maximum grain sizes from the curves in Figure 4 of \citet{Kataoka2017}, we find all of the above values of the total polarization fraction (at their respective wavelengths) are roughly consistent with a maximum size of $\sim$\,70\,$\micron$ for the grains producing the polarized emission.  These polarization fractions (calculated by averaging across the entire disk) are consistent with the previous non-detections of polarization in lower-resolution (1--2$\arcsec$) observations of Class II circumstellar disks \citep{Hughes2009b, Hughes2013}; the detection of 870\,$\micron$ polarization in both IM~Lup and HL~Tau suggests that the polarized emission produced by self-scattering might have been detectable in the aforementioned CARMA and SMA observations if they had had the sensitivity to detect polarization fractions at the $\lesssim$\,0.5\% level.



\bigskip
On one hand, the fact that a variety of distinct models yield similar maximum grain-size estimates for HL~Tau and IM~Lup is encouraging.
On the other hand, considering the major differences between the two disks (e.g., both polarization fraction and evolutionary state), the fact that the maximum grain sizes for both HL~Tau and IM~Lup are $\sim$\,70\,$\micron$ is unexpected.  Furthermore, it is unnerving that $\sim$\,70\,$\micron$ is so different from the millimeter/centimeter sizes of grains that many studies have inferred to be responsible for the bulk of the detected (unpolarized) dust emission in disks \citep{Perez2012, Trotta2013, Perez2015, Tazzari2016, YLiu2017}.  This issue is discussed in \citet{Kataoka2016} and \citet{Yang2016a}, where they mention several possible ways to reconcile the maximum grain-size estimates from scattering models and from models of the dust emission spectrum: for example, scattering by ``fluffy,'' porous grains as opposed to the simple spherical grains assumed by Mie theory; or separate populations of grains responsible for the observed polarization versus the bulk of the unpolarized emission.  

\subsection{Constraining dust settling using disk near/far-side asymmetry and polarization orientations}
\label{sec:dust_settling}

\smallskip
\subsubsection{Scattering and optical depth}

Polarization from self-scattering is most easily produced when the disk is moderately optically thick.  When discussing optical depth in the context of dust self-scattering models, one must consider two optical depths for the dust layer: one perpendicular to the disk plane, $\tau_\perp$, and the other through the mid-plane of the disk, $\tau_\parallel$. The one typically referred to is $\tau_\perp$, which is essentially the column density, and which controls the percentage of the scattering-produced polarization for given grain properties.  Scattering models suggest that the maximum polarization fraction occurs near $\tau_\perp \approx 1$ \citep{Yang2017}; to have an appreciable polarization, the optical depth $\tau_\perp$ cannot be \textit{too} small, but the actual number will depend on grain properties.  

\citet{Cleeves2016} find that the inner $\sim$\,40\,au of the disk is optically thick ($\tau_\perp \gg 1$).  Their model finds that the rest of the disk is optically thin ($\tau_\perp \lesssim 1$); however, they also report tentative evidence of ringed structures, which could potentially be optically thick, but are not resolved at the low resolution of their (or our) observations.  Based on results from \citet{Stephens2017b}, the ringed structures in HL~Tau have sufficient optical depth to produce scattering; the effect of such rings on disk polarization profiles was discussed by \citet{Kataoka2016} and \citet{Pohl2016}.  Future modeling will allow us to explore whether concentrating (sub)millimeter-sized grains into rings (as expected by pressure-maximum-induced dust trapping) would make dust self-scattering more or less efficient, and whether the polarization pattern would change.  

\subsubsection{Constraining dust settling in IM~Lup}

Based on their modeling of ALMA 1.3\,mm and 870\,$\micron$ continuum and spectral-line observations toward IM~Lup, \citet{Cleeves2016} found that the gas and the micron-sized grains in the IM~Lup disk have a significant vertical extent (with a scale height of 12\,au at a radius of 100\,au; this is consistent with Very Large Telescope (VLT) SPHERE/IRDIS observations of IM~Lup in optical scattered light; \citealt{Avenhaus2018}).  The scale height of the millimeter-sized grains is $\sim$\,4\,$\times$ less; however, to reproduce the significant depletion of spectral-line emission in the inner $\sim$\,40\,au of the disk (which they attribute to absorption by optically thick dust), some amount of turbulent lofting of millimeter grains is necessary. 

Dust settling can be constrained with scattering models (e.g., \citealt{Kataoka2015, Yang2017}).  \citet{Yang2017} have shown that, in inclined disks such as IM~Lup, the near side of the disk would be significantly brighter in polarized intensity than the far side if the disk were optically thick (i.e., $\tau_\perp > 1$) \textit{and} the grains responsible for the scattering-induced polarization had a significant vertical extent. There is no obvious near/far-side asymmetry detected in our observations (Figure \ref{fig:obs}); thus, if IM~Lup were marginally optically thick ($\tau_\perp \approx 1$) in the region where polarization is detected, it would be reasonable to conclude that the $\sim$\,60\,$\micron$-sized grains that efficiently scatter the 870\,$\mu$m photons have already settled into a relatively thin layer (this is similar to what was found in HL~Tau by \citealt{Stephens2017b}).  This settling may be inconsistent with the lofting of large grains\footnote{Here we categorize both the $\sim$\,60\,$\micron$ grains in IM~Lup and the millimeter-sized grains in the \citet{Cleeves2016} models as ``large.''} expected based on the work by \citet{Cleeves2016}; however, based on their modeling, it is not clear what the exact optical depth is across the region where we detect polarization, and thus we cannot currently draw firm quantitative conclusions about dust settling.

At larger radii, where the disk is optically thin in the vertical direction (i.e., $\tau_\perp <1$), one can in principle constrain the geometric thickness of the dust layer using the polarization orientation relative to the (projected) disk minor axis. The reason is that the orientation of the scattering-induced polarization at a given location in an inclined disk depends on the anisotropy of the incident radiation field in the plane of the disk at that location. If the disk-plane radiation field is nearly isotropic, as would be the case if the optical depth in the plane of the disk were larger than unity (i.e., $\tau_\parallel \gtrsim 1$), the polarization of the scattered light would be along the minor axis of the inclined disk. If, on the other hand, the dust were optically thin in the disk plane, the incident radiation field would be strongly beamed in the radial direction, which would lead to a more azimuthal polarization orientation. Since most of the polarization orientations observed in IM~Lup are along the minor axis (see Figure \ref{fig:obs}), it is likely that the region where significant polarization is detected is optically thick along the disk plane (i.e., $\tau_\parallel \gtrsim 1$). Since $\tau_\parallel$ depends on the geometric thickness, $H$, of the disk (i.e., for a given surface density, the smaller $H$ is, the larger $\tau_\parallel$ would become), polarization observations---in concert with full radiative-transfer modeling---could in principle be used to constrain the thickness of the layer of grains in IM~Lup that are responsible for the observed polarization pattern.

\subsection{Magnetic fields or scattering in the IM~Lup disk?}
\label{sec:disk_Bfields}

Magnetic fields are thought to play a crucial role in the dynamics and evolution of disks around young stars via the MRI and magnetized disk winds.  If a disk were fully turbulent due to the MRI, the magnetic field orientation would change on dynamical timescales, and would be dominated by the turbulent component of the magnetic field.  In this case, grain alignment would be possible only when it occurs very quickly, and the resulting dust-polarization configuration would either exhibit random patterns or would be unpolarized. However, observational and theoretical studies point toward weak turbulence \citep{Hughes2011, Guilloteau2012, Flaherty2015, Simon2015, Teague2016, Flaherty2017, Flaherty2018} and largely laminar magnetic fields with more stable magnetic structures dominated by a toroidal field component \citep[e.g.,][]{Bai2017}. In this case, alignment of grains with the magnetic field is possible only when the alignment timescale is shorter than the gas damping timescale. \citet{Tazaki2017} showed that the precession timescale of millimeter-sized grains around magnetic fields (which is the first step toward magnetic grain alignment) can significantly exceed the gas damping timescale in the outer disk, making alignment unfavorable in that region.  Nevertheless, the field strength adopted by \citet{Tazaki2017} is relatively low (the field strength they adopt is equal to their Equation 49 multiplied by 10, and is $\sim$\,100\,$\mu$G in the outer disk).

The magnetic alignment timescale can be brought down in the presence of a stronger magnetic field (and hence a larger accretion rate, e.g., Equations 7 and 16 of \citealt{Bai2009}).  Recent estimates of IM~Lup's accretion rate suggest a rate of $10^{-8}$\,$M_\odot$\,yr$^{-1}$ \citep{Alcala2017}, a typical value for Class II disks.  If accretion is not far from steady state in IM~Lup, it implies a $\sim$\,1--5\,mG field strength at the $\sim$\,50--100\,au scale.\footnote{The steady-state assumption in this estimate may not hold. Recent theoretical work by \citet{Simon2017} suggests that in order to have turbulent line-widths consistent with the current observational constraints, the magnetic field strength in the outer disk is likely to be very weak (e.g., $\sim$\,20\,$\mu$G at 100\,au in the disk of HD 163296, analyzed by \citealt{Flaherty2015, Flaherty2017}), making the accretion rate there much smaller than the instantaneous accretion rate.}
Considering the calculations by \citet{Tazaki2017}, such a field strength still appears insufficient to produce alignment in millimeter-sized grains.  On the other hand, alignment might still be possible for Class 0/I disks with higher accretion rates (and hence stronger fields), and there is indeed evidence of magnetically aligned grains in the very early stages of star formation \citep{Cox2015, SeguraCox2015, Liu2016}.

If the dominant grain size is $\sim$\,60\,$\micron$, as we infer above, then the precession timescale around the magnetic field is comparable to or smaller than the gas damping timescale, assuming the grains contain super-paramagnetic inclusions (whose properties are very uncertain). The magnetic alignment timescale is expected to be larger than the precession timescale by an unknown factor (\citealt{Tazaki2017} assume a factor of 30). Given the substantial uncertainties, the case for magnetic alignment of $\sim$\,60\,$\micron$ grains is unclear, but may be possible in the outer disk of IM~Lup (where we detect no polarization).  On the other hand, magnetic alignment becomes more favorable for smaller grains ($\sim$\,1--10\,$\micron$ in size), as was discussed in \citet{DLi2016}.

It is also worth noting that based on \citet{Tazaki2017}, it appears that the timescales for alignment of both $\sim$\,60\,$\micron$ and sub-millimeter-sized grains by the anisotropic radiation field are well below both the gas damping timescale and the magnetic alignment timescale, thus favoring alignment with the radiation field. However, we do not see an obvious contribution to the polarization pattern from alignment with the radiation field, suggesting that there are still substantial uncertainties in our understandings of the grain alignment theory.

\section{Conclusions}
\label{sec:con}

We have presented 870\,$\micron$ ALMA observations of polarized dust emission toward the protoplanetary disk surrounding the Class~II source IM~Lup.  After analyzing the IM~Lup polarization maps and comparing the results with previously published polarization data toward the Class I/II source HL~Tau, we come to the following conclusions:

\begin{enumerate}

\item We find that the orientation of the polarized emission is along the minor axis of the disk, and that the value of the polarization fraction increases steadily toward the center of the disk, reaching a peak value of $\sim$\,1.1\%.  All of these characteristics are consistent with models of self-scattering of submillimeter-wave emission from an optically thin inclined disk.

\item The distribution of the polarization position angles across the IM~Lup disk reveals that while the average orientation is along the minor axis, the polarization orientations show a significant spread in angles.  This is consistent with models of pure scattering.

\item We compare the polarization of IM~Lup with that of HL~Tau.  A comparison of cuts of the polarization fraction across the major and minor axes of both sources reveals that IM~Lup has a substantially higher polarization fraction than HL~Tau toward the center of the disk.  This enhanced polarization fraction could be due a number of factors, including higher optical depth in HL~Tau, or scattering by larger dust grains in the more evolved IM~Lup disk.  However, models yield similar maximum grain sizes for both HL~Tau (72\,$\micron$) and IM~Lup (61\,$\micron$, this work).  This reveals continued tension between grain-size estimates from scattering models and from models of the dust emission spectrum, which find that the bulk of the (unpolarized) emission in disks is most likely due to millimeter (or even centimeter) sized grains. 

\end{enumerate}

The mounting evidence for scattering in (sub)millimeter-wavelength observations of protoplanetary disks, combined with the evidence for different polarization mechanisms at different wavelengths \citep{Kataoka2017, Stephens2017b}, yield several requirements for progress in the field of disk polarization.  The first is multi-wavelength observations, which, in combination with models and synthetic observations, will allow us to disentangle the various mechanisms that may be causing disk polarization, including dust self-scattering, alignment with the dust emission gradient, and alignment with the magnetic field.  The second is higher-resolution observations of polarization in disks (given sufficient signal-to-noise) in order to understand the effects of disk sub-structure such as rings \citep[e.g.,][]{ALMA2015} and spiral arms \citep{Perez2016,Boehler2018} on polarization profiles.  

Finally, in light of the scant evidence for magnetically aligned grains at (sub)millimeter wavelengths in Class II disks, it seems likely that the best way to directly detect magnetic fields in disks will be via the Zeeman effect \citep{Brauer2017} or via spectral-line polarization (e.g., the Goldreich-Kylafis effect; \citealt{Goldreich1981}).  Spectral-line polarization observations are already possible with ALMA; Zeeman observations will also soon be possible (although \citeauthor{Brauer2017} note that the signal will be difficult to detect in all but the brightest disks with high [$\gtrsim$\,1\,mG] magnetic field strengths and high abundances of CN.)  These spectropolarimetric observations will allow us to continue to refine our understanding of the role of the magnetic in star formation at the scales of protoplanetary disks, while in parallel we can use observations of scattering-induced polarization to probe grain growth in protoplanetary disks, which is a crucial step toward the formation of planets \citep{Testi2014}.

\begin{table*}[hbt!]
\small
\begin{center}
\caption{\normalsize \vspace{0.1in} ALMA polarization data}
\begin{tabular}{cccccccc}
\hline
\hline
$\alpha_{\textrm{J2000}}$ & $\delta_{\textrm{J2000}}$ & $I$ & $P$ & $\chi$ & $\delta\chi$ & $P_{\textrm{frac}}$ & $\delta P_{\textrm{frac}}$ \\
(\degree) & (\degree) & \mjybmvert & \mjybmvert & (\degree) & (\degree) & & \vspace{0.25em} \\
\hline
 239.03854 & --37.93539 &   15.953 &   0.086 &    51.7 &     6.2 &   0.0054 &   0.0011 \\ 
 239.03846 & --37.93539 &   18.887 &   0.112 &    48.3 &     4.7 &   0.0059 &   0.0010 \\ 
 239.03839 & --37.93539 &   19.832 &   0.106 &    50.6 &     4.9 &   0.0053 &   0.0009 \\ 
 239.03832 & --37.93539 &   18.730 &   0.076 &    66.4 &     6.8 &   0.0041 &   0.0010 \\ 
 239.03854 & --37.93533 &   19.116 &   0.111 &    40.5 &     4.7 &   0.0058 &   0.0009 \\ 
 239.03846 & --37.93533 &   24.032 &   0.164 &    43.4 &     3.2 &   0.0068 &   0.0007 \\ 
 239.03839 & --37.93533 &   27.408 &   0.188 &    50.9 &     2.8 &   0.0068 &   0.0007 \\ 
 239.03832 & --37.93533 &   27.130 &   0.163 &    59.1 &     3.2 &   0.0060 &   0.0007 \\ 
 239.03824 & --37.93533 &   23.048 &   0.111 &    55.7 &     4.7 &   0.0048 &   0.0008 \\ 
 239.03817 & --37.93533 &   17.343 &   0.069 &    45.8 &     7.5 &   0.0040 &   0.0010 \\ 
 239.03861 & --37.93527 &   15.020 &   0.079 &    23.8 &     6.6 &   0.0052 &   0.0012 \\ 
 239.03854 & --37.93527 &   21.742 &   0.134 &    32.5 &     3.9 &   0.0062 &   0.0008 \\ 
 239.03846 & --37.93527 &   31.406 &   0.242 &    43.5 &     2.1 &   0.0077 &   0.0006 \\ 
 239.03839 & --37.93527 &   43.068 &   0.369 &    49.7 &     1.4 &   0.0086 &   0.0004 \\ 
 239.03832 & --37.93527 &   48.296 &   0.403 &    51.2 &     1.3 &   0.0084 &   0.0004 \\ 
 239.03824 & --37.93527 &   39.648 &   0.284 &    48.5 &     1.8 &   0.0072 &   0.0005 \\ 
 239.03817 & --37.93527 &   25.788 &   0.105 &    49.0 &     5.0 &   0.0041 &   0.0007 \\ 
 239.03854 & --37.93521 &   22.950 &   0.101 &    35.2 &     5.1 &   0.0044 &   0.0008 \\ 
 239.03846 & --37.93521 &   39.601 &   0.293 &    46.6 &     1.8 &   0.0074 &   0.0005 \\ 
 239.03839 & --37.93521 &   71.463 &   0.733 &    49.9 &     0.7 &   0.0103 &   0.0003 \\ 
 239.03832 & --37.93521 &   96.251 &   1.069 &    50.3 &     0.5 &   0.0111 &   0.0002 \\ 
 239.03824 & --37.93521 &   76.670 &   0.780 &    49.2 &     0.7 &   0.0102 &   0.0002 \\ 
 239.03817 & --37.93521 &   41.019 &   0.283 &    47.7 &     1.8 &   0.0069 &   0.0004 \\ 
 239.03854 & --37.93515 &   21.263 &   0.084 &    41.6 &     6.3 &   0.0039 &   0.0008 \\ 
 239.03846 & --37.93515 &   40.420 &   0.279 &    49.2 &     1.9 &   0.0069 &   0.0004 \\ 
 239.03839 & --37.93515 &   83.335 &   0.835 &    52.0 &     0.6 &   0.0100 &   0.0002 \\ 
 239.03832 & --37.93515 &  118.774\phantom{1} &   1.301 &    52.7 &     0.4 &   0.0110 &   0.0002 \\ 
 239.03824 & --37.93515 &   94.140 &   0.974 &    52.3 &     0.5 &   0.0103 &   0.0002 \\ 
 239.03817 & --37.93515 &   49.330 &   0.405 &    50.0 &     1.3 &   0.0082 &   0.0004 \\ 
 239.03809 & --37.93515 &   25.967 &   0.113 &    42.1 &     4.6 &   0.0043 &   0.0007 \\ 
 239.03854 & --37.93510 &   17.115 &   0.069 &    44.7 &     7.7 &   0.0041 &   0.0011 \\ 
 239.03846 & --37.93510 &   28.915 &   0.160 &    45.3 &     3.2 &   0.0055 &   0.0006 \\ 
 239.03839 & --37.93510 &   51.895 &   0.414 &    52.8 &     1.2 &   0.0080 &   0.0003 \\ 
 239.03832 & --37.93510 &   70.911 &   0.643 &    55.3 &     0.8 &   0.0091 &   0.0003 \\ 
 239.03824 & --37.93510 &   62.504 &   0.597 &    54.7 &     0.9 &   0.0096 &   0.0003 \\ 
 239.03817 & --37.93510 &   41.216 &   0.359 &    51.2 &     1.4 &   0.0087 &   0.0004 \\ 
 239.03809 & --37.93510 &   25.989 &   0.150 &    43.3 &     3.5 &   0.0058 &   0.0007 \\ 
 239.03839 & --37.93504 &   27.455 &   0.133 &    59.3 &     3.9 &   0.0048 &   0.0007 \\ 
 239.03832 & --37.93504 &   34.876 &   0.248 &    60.4 &     2.1 &   0.0071 &   0.0005 \\ 
 239.03824 & --37.93504 &   35.688 &   0.300 &    55.6 &     1.7 &   0.0084 &   0.0005 \\ 
 239.03817 & --37.93504 &   30.041 &   0.238 &    50.3 &     2.2 &   0.0079 &   0.0006 \\ 
 239.03809 & --37.93504 &   22.773 &   0.161 &    42.3 &     3.2 &   0.0071 &   0.0008 \\ 
 239.03802 & --37.93504 &   16.565 &   0.094 &    38.4 &     5.6 &   0.0057 &   0.0011 \\ 
 239.03832 & --37.93498 &   21.929 &   0.117 &    55.5 &     4.4 &   0.0054 &   0.0008 \\ 
 239.03824 & --37.93498 &   23.556 &   0.141 &    52.2 &     3.7 &   0.0060 &   0.0008 \\ 
 239.03817 & --37.93498 &   22.219 &   0.120 &    46.5 &     4.3 &   0.0054 &   0.0008 \\ 
 239.03809 & --37.93498 &   18.752 &   0.115 &    41.7 &     4.5 &   0.0062 &   0.0010 \\ 
 239.03802 & --37.93498 &   14.755 &   0.079 &    37.5 &     6.7 &   0.0053 &   0.0012 \\ 
 239.03824 & --37.93492 &   16.979 &   0.077 &    48.9 &     6.7 &   0.0045 &   0.0011 \\ 
 239.03817 & --37.93492 &   16.740 &   0.068 &    42.9 &     7.5 &   0.0041 &   0.0011 \\ 
 239.03824 & --37.93486 &   11.755 &   0.068 &    42.8 &     7.6 &   0.0058 &   0.0015 \\ 
 239.03817 & --37.93486 &   11.965 &   0.076 &    43.5 &     6.9 &   0.0064 &   0.0015 \\ 

\hline
\vspace{-1.5em}
\end{tabular}
\label{table:data}
\end{center}
\footnotesize
\textbf{Note.}  $I$ is the total intensity, reported where $I > 3\,\sigma_I$.  $P$ is the polarized intensity, reported where $P > 3\,\sigma_P$. $\chi$ is the orientation of the polarization, measured counterclockwise from north.  $\delta\chi$ is the statistical uncertainty in the polarization orientation.  $P_{\textrm{frac}}$ is the polarization fraction $P/I$, reported where $P > 3\,\sigma_P$.  $\delta P_{\textrm{frac}}$ is the statistical uncertainty in the polarization fraction (note that in the highest signal-to-noise points, these values are smaller than the 0.03\% systematic uncertainty in $P_{\textrm{frac}}$).
\textit{This table, in machine-readable format, is available in the online version of this publication.}  
\vspace{1em}
\end{table*}

\acknowledgments
The authors thank the anonymous referee, whose comments led to substantial improvements in the manuscript.
C.L.H.H. acknowledges the calibration and imaging work performed at the North American ALMA Science Center.
H.Y. is supported in part by an SOS award from NRAO. 
Z.-Y.L. is supported in part by NASA NNX 14AB38G and NSF AST-1313083 and 1716259.
L.I.C. acknowledges the support of NASA through Hubble Fellowship grant HST-HF2-51356.001-A awarded by the Space Telescope Science Institute, which is operated by the Association of Universities for Research in Astronomy, Inc., for NASA, under contract NAS 5-26555.
The authors acknowledge Karin \"Oberg for providing continuum maps of IM~Lup during the ALMA proposal process, and Richard Teague for the helpful discussion about upper limits on turbulence in protoplanetary disks. 
This paper makes use of the following ALMA data: ADS/JAO.ALMA\#2016.1.00712.S.
ALMA is a partnership of ESO (representing its member states), NSF (USA) and NINS (Japan), together with NRC (Canada), MOST and ASIAA (Taiwan), and KASI (Republic of Korea), in cooperation with the Republic of Chile. The Joint ALMA Observatory is operated by ESO, AUI/NRAO and NAOJ.
The National Radio Astronomy Observatory is a facility of the National Science Foundation operated under cooperative agreement by Associated Universities, Inc. 
This research made use of APLpy, an open-source plotting package for Python hosted at \url{http://aplpy.github.com}.

\textit{Facilities:} ALMA.

\bibliography{ms}
\bibliographystyle{apj}

\end{document}